\documentclass[oneside,english]{amsart}
\usepackage[T1]{fontenc}
\usepackage[latin9]{inputenc}
\usepackage{geometry}
\usepackage{textcomp}
\usepackage{amstext}
\usepackage{amsthm}

\makeatletter
\numberwithin{equation}{section}
\numberwithin{figure}{section}

\makeatother

\usepackage{babel}
\begin{document}
\title{On something and nothing: The interface of two types of nothing}
\author{Adam Brownstein$^{*}$}
\thanks{$^{*}$University of Queensland, Australia. ORCID: https://orcid.org/0009-0001-7814-4384}
\begin{abstract}
We suggest that the question of why is there something rather than
nothing can be answered by the existence of two types of nothing.
We propose that matter occurs at the boundaries of intersection of
both nothings. This accords with the common view that there are three
worlds; the platonic world of concepts, the material world (i.e. of
matter), and the non-material world (i.e. of consciousness). Both
the material and non-material worlds have their own type of nothing,
thus leading to the proposal. The interpretation provides an alternative
type of duality distinct from property or substance dualism, and may
unify the two. The interpretation also has implications for the understanding
of physical causation, and Mach's principle as the boundary of intersection
provides a return to the aether concept. 
\end{abstract}

\maketitle

\section{Introduction}

\noindent The question of why there is something rather than nothing
is a foundational question in metaphysics. The question is why is
there not simply a void, with no matter, no consciousness, and no
laws of physics. In terms of pure logic, the question has no possible
answer. It is not conceivable to have something arising out of a single,
absolute nothing. One must make primitive assumptions about what nothing
consists of to attain an answer; otherwise the universe must be eternal,
perhaps existing in a cyclical Big-Bang cosmology. 

For instance, one can take as a base assumption the laws of quantum
mechanics and general relativity. Therefore the state of nothing is
conceived of as a quantum vacuum state. Under this assumption, the
instability of the quantum vacuum state, processes of quantum field
theory, and cosmic inflation together can be used to describe the
generation of something from nothing. We suggest that this line of
thinking, while highly informative from a scientific view, is erroneous
in assuming the laws of physics hold true for a state of the universe
containing no matter. It is conceivable that the laws of physics arise
out of the properties of the matter present in the space, rather than
being external and nomological or law-like. 

Because the quantum vacuum state has physical laws, it is likely not
physically empty as these laws derive from the matter present in the
space. The quantum vacuum may be a highly active foam of virtual particles
and quantum fields, which have a physical reality in the realist ontologies
of the wavefunction or quantum fields. 

Is it possible to simplify the assumptions, and propose a less restrictive
form of nothing than the quantum vacuum state? A state which assumes
no prior laws of physics, and yet either implies the creation of the
universe, or the impossibility of its non-existence? We argue that
it is possible to achieve this by assuming the necessary existence
of a blank physical space, containing no matter, no laws of physics
but material and non-material nothing. If there are two types of nothing,
then this may imply the generation of something from nothing in a
primitive way. 

If it is assumed that the material and non-material nothing are mutually
exclusive, yet share the same physical space, then the result is obtained.
It remains to be proven that the material and non-material nothing
are a logical necessity, given that they are known about in our universe. 

\subsection*{Occam's razor }

In light of this perspective, there are three main alternative hypotheses: 
\begin{enumerate}
\item The universe containing matter exists for no reason. It may have existed
eternally (e.g. in a cyclical Big-Bang cosmology), or spontaneously
arisen for the first time during the Big-Bang.
\item There were initial states which existed prior, for instance a quantum
vacuum state; and these initial states themselves either existed eternally
or arose spontaneously. 
\item The material and non-material worlds alone existed prior. Physical
states of matter formed out of the boundary of intersection of two
types of nothing. 
\end{enumerate}
We suggest that simplest explanation is the third. The first explanation
is unsatisfactory as it cannot answer the question in principle. The
second answer makes more assumptions than the third answer, as to
have an initial physical state this already implies the existence
of the material and non-material worlds, which alone is sufficient
for the third answer. Using the third answer, it is also potentially
possible to explain how the initial physical conditions such as the
quantum vacuum state themselves arose, therefore it precedes the second
answer.

An entirely different reason to prefer the third answer is that it
provides a means to understand dualism. The model of two intersecting
types of nothing unifies property and substance dualism. In doing
so, it removes the gap between the material and non-material worlds,
thus enabling a more comprehensive picture of physical causation involving
both the material and non-material aspects. 

One might be concerned that replacing the something of physical matter
or consciousness with nothing does not accurately represent its nature
as a substance. However, it is clear that experimentally the proposed
something inside matter can never directly be accessed. It is always
observed in a relational capacity. This is the inherent simplicity
of the model; it recognizes that substance as a definite object is
redundant ontologically, and can never be empirically observed. Therefore,
why not replace it with nothing? Nothing is lost by making this replacement,
and one gains a picture of reality closer to that which can be empirically
ascertained. The boundary of intersection of two types of nothing
can adequately describe the relational aspect of physical matter,
without requiring an inner substance. 

\section{The existence of three worlds}

\subsection{Hard problem of consciousness}

\noindent A hint toward this solution comes from the hard problem
of consciousness. It is shown through logical argument; for instance
the philosophical zombie argument \cite{key-2}; that consciousness
is distinct from the material world. Therefore, a pure materialist
viewpoint is evidently false, and one might take either the idealist
approach (everything is non-material) or dualist approach (physical
reality has material and non-material aspects).

Taking the dualist view, if there are two types of something (material
and non-material something), then there may also be two types of nothing
(material and non-material nothing). This understanding opens up new
possible approaches to the question of why there is something rather
than nothing. 

\subsection{Three worlds}

In the dualist interpretation there are three worlds of existence:
\begin{enumerate}
\item \emph{Platonic world: }Including logic, mathematical relationships,
geometric forms, and truth statements. Mathematics is the investigation
of the structures present in this world. 
\item \emph{Material world:} This is the world of atoms, particles, wavefunctions,
and material substances. It is assumed according to the realist view
that there is physical matter which obeys the equations of physics
and occurs in the material world. 
\item \emph{Non-material world: }The hard problem of consciousness establishes
that consciousness cannot be part of the material world. Therefore,
there is a non-material world (or non-material phenomena or properties)
separate from the material world.
\end{enumerate}
The existence of these three worlds is explained clearly in works
such as that of Roger Penrose \cite{key-1}. Although these worlds
have been known since antiquity, for instance in Descartes's dualism,
and the Ancient Greek understanding of mind, body and platonic forms. 

Note that although we have described three worlds, they are not necessarily
separate. The material and non-material worlds can exist inside the
same physical space in some models of dualism. For instance, in property
dualism, matter has a material and non-material aspect, and therefore
the material and non-material worlds coincide in the same physical
space. 

\section{The proposed solution}

\subsection{The solution in a nutshell }

As we have established, if one takes the hard problem of consciousness
seriously, then the possibility that there are two types of nothing
becomes immediately evident. Just as there is the possibility for
a material something and non-material something, there is an equal
possibility for a material nothing and non-material nothing.

What then do two types of nothing look like? Firstly, one might envisage
that this is just empty space. In this case, the nothing of the material
and non-material worlds together generate the traditional conception
of nothing. However, this interpretation assumes either of two things;
firstly, that the material and non-material worlds are separate and
do not share the same physical space; alternatively, if they do share
the same physical space, then they overlap in spatial location as
would be the case in standard interpretations of property dualism. 

We reject these assumptions, especially in light of substance dualism.
If there are two substances, it is plausible that the material and
non-material substances share the same physical space but may exclude
one another, and compete with one another to occupy the spatial locations.
If we suppose that the material nothing and non-material nothing occupy
the same physical space, yet cannot overlap in spatial location, then
they must form an interface between them; unless there is a third
type of nothing in between them (which there is no present evidence
for). 

Furthermore, if one type of nothing dominates the whole space, then
the other type of nothing will not be present, which we assume to
be a contradiction (which is an important point to be discussed in
further detail later). Therefore, the boundary between both nothings
is always implied. We hypothesize that physical matter really is this
boundary of intersection between both nothings. 

\subsection{Old model: Single nothing}

In the old model, there is a single type of nothing. Therefore, the
basic configuration of the universe one of the following possibilities: 
\begin{enumerate}
\item (Something 1, Something 2): Dualism
\item (Something 1, Nothing): Materialism
\item (Nothing, Something 2): Idealism 
\item (Nothing, Nothing): Empty physical space
\end{enumerate}
In this categorization, the first position corresponds to the material
world, and the second position to the non-material world. There is
also an additional possibility which has been omitted: 
\begin{enumerate}
\item (No nothing \& No something 1, No nothing \& No something 2)
\end{enumerate}
which is the abstract total void state. 

The problem with this model is that even if we are able to argue that
the abstract total void is logically impossible, the case of empty
physical space, namely the state of (Nothing, Nothing), is always
a possibility. This prevents the proof of nothing conceived of as
empty physical space being logically impossible, and so no real progress
can be made, aside from introducing either material or non-material
substances from which the early universe can arise.

\subsection{New model: Two types of nothing }

If instead there are two types of nothing, then the possible states
of the universe are: 
\begin{enumerate}
\item (Something 1, Something 2): Dualism 
\item (Something 1, Nothing 2): Materialism
\item (Nothing 1, Something 2): Idealism
\item (Nothing 1, Nothing 2): Either dualism, or empty physical space; depending
on whether the two types of nothing i) share the same physical space
and ii) are non-overlapping or overlapping in this space. 
\end{enumerate}
There are also five possible void states: 
\begin{enumerate}
\item (Something 1, $\text{No nothing 2 }\&\text{ No something 2}$): Materialism
\item (Nothing 1, $\text{No nothing 2 }\&\text{ No something 2}$): Non-material
void, empty material space
\item ($\text{No nothing 1 }\&\text{ No something 1}$, Something 2): Idealism 
\item ($\text{No nothing 1 }\&\text{ No something 1}$, Nothing 2): Material
void, empty non-material space
\item ($\text{No nothing 1 }\&\text{ No something 1}$, $\text{No nothing 2 }\&\text{ No something 2}$):
Abstract total void
\end{enumerate}
Taking the dualist viewpoint excludes cases 1 and 3, which are expressions
of materialism and idealism respectively. Cases 2, 4 and 5 are excluded
on the basis that void states are impossible, as we assume the existence
of the material and non-material worlds either as a brute fact or
a logical necessity. 

To explain the void states, having the combination no nothing 1, and
no something 1 indicates that the material world doesn't exist, which
violates the assumption that the material world is present. Likewise,
having both no nothing 2 and no something 2 indicates that the non-material
world doesn't exist, which violates the assumption that the non-material
world is present. 

\subsection{Complexity of the boundary}

Assuming the existence of the bare material and non-material worlds,
the question of why there is something rather than nothing can be
understood to a degree. However, the question shifts to explaining
why the boundary of intersection between both types of nothing is
interesting. Why does it form complexity, rather than being a flat
sheet of intersection between the worlds? 

There are a range of questions to consider, which may indicate how
the boundary could compose itself into structured physical matter.
For example, what is the spatial extent of the boundary of intersection?
Is it stable in configuration or unstable? 

To answer why the interface might form complexity, we highlight that
a space filled with two types of nothing may have an arbitrary arrangement.
It may be inherently unstable, with no reason to prefer one configuration
over another. There is no reason, for instance, to prefer the configuration
of an infinite flat sheet (or locally flat sheet which conforms to
the geometry of the universe, if the universe is closed). In asking
why is there complexity, we are assuming that a flat sheet is the
preferred option over the infinitely many other arrangements which
have complexity, which may be an erroneous assumption to make. So
it may simply be a matter of arbitrariness why complexity is present.

In relation to arbitrariness, the movement of the boundary is constrained
by the requirement that it remains a continuous structure. Yet each
location along the boundary has an element of uncertainty due to the
inherent arbitrariness. Together this may cause entropic motion, while
preserving the structure of the boundary. Therefore, the movement
and formation of complex structure in the boundary may be an entropic
phenomenon, not necessarily due to physical forces. 

Secondly, there may be a type of pressure exerted by different arrangements
of the boundary. For instance, suppose a closed universe. If the boundary
of intersection is also a closed surface to match this geometry, then
there is an interior and exterior. Having a closed interior may create
a surface pressure on the sheet. A constantly changing surface pressure
may imply the sheet has motion and forms complexity. 

In any case, there are plausible mechanisms for the formation of complexity
within the boundary of intersection. This is far superior than attempting
to explain how something can arise out of the empty space in the case
of a single nothing. The boundary of intersection acts as a base type
of substance or field from which physical matter can emerge. If there
is instead a single type of nothing, all that is present is a blank
physical space with no possibility for matter to emerge spontaneously
from the space. 

\section{Assumptions and discussion}

\noindent We have assumed in this proposal that nothing means the
bare physical space, containing the empty material and non-material
worlds. We have assumed that the material and non-material worlds
exist even if they contain nothing. This assumption requires further
thought, as it touches upon deep philosophical questions such as whether
it is possible to conceive of void states.

\subsection{Ansatz}

We have assumed that the two types of nothing occupy the same physical
space, yet are mutually exclusive in spatial location. This is not
an assumption requiring direct proof, it is a model or ansatz that
is used for illustration of the hypothesis. We clarify this point
so that attention can be paid to the true assumptions of our proposal,
the existence of the material and non-material worlds and the philosophy
of void-states. 

\subsection{Assumptions }
\begin{enumerate}
\item \emph{Existence of a bare material world: }This is simple to understand
by the hypothetical removal of all physical matter from space. This
leaves the empty material world which is the blank physical space. 
\item \emph{Existence of a bare non-material world:} We suggest that the
non-material world also must continue to exist even if it contains
nothing. Just as the material world continues to exist even if it
contains nothing, by analogy we suggest the same is true of the non-material
world. 
\item \emph{Void states are logically impossible: }Although we are not completely
certain on the validity of this assumption, we are able to make somewhat
compelling arguments (e.g. the principle of indifference) that the
void states are logically impossible given that the material and non-material
worlds are known to exist. These void states are i) the material void
ii) the non-material void and iii) the abstract total void of material
void and non-material void combined. 
\end{enumerate}

\subsection{Principle of indifference}

To put our argument for the impossibility of void states succinctly,
it can be reduced down to the following principle: Once a particular
form of nothing (e.g. background space, material nothing, non-material
nothing) is learned to exist in a single universe, then the background
space or world containing this nothing must necessarily exist in all
possible universes. 

This follows because it is always conceivable to have the universe
containing this nothing, which is a functionally equivalent state
to having the universe absent the nothing. There is an indifference
between the two situations, and this indifference may imply it is
impossible to ever remove the concept in principle. Therefore, once
a type of nothing is discovered to exist, it becomes a brute fact. 

This is similar to how a mathematical truth in the platonic world
necessarily exists in all possible universes, once it is learned to
exist in a single universe. Elements of the platonic world might not
be the only pieces of knowledge with this property; knowledge about
the basic physical structure of the universe; such as the existence
of the material and non-material worlds, and the dimensionality of
the space; may likewise have this property. 

\subsection{Further discussion}

We acknowledge that the principle of indifference is not provable
to be true, and that void states are conceivable ontologically. However,
we do believe the principle is worth consideration. 

For instance, it might not be conceivable to have a two-dimensional
universe now that we understand the logical possibility of a third
dimension. This situation follows from the principle of indifference.
Any two dimensional universe is identical to having the third-dimension
filled with nothing. So once the third dimension is known about, may
be impossible to truly remove it as a concept. This is especially
true if space is viewed in a relational sense, rather than as a concrete
reality. 

\section{\label{sec:Dualism}Dualism}

\noindent The viewpoint of two mutually exclusive types of nothing
forming an interface may provide a more coherent understanding of
dualism. This is an important piece of evidence supporting the model,
because it appears to provide a comprehensive picture of dualism,
which makes sense of many of the known problems of dualism. 

\subsection{Addressing the problems of dualism }

Dualism has some known ontological issues which are worth revisiting
in light of the new model: 
\begin{enumerate}
\item \emph{Property dualism: }Property dualism lends itself to taking a
panpscyhist view that all matter has a non-material aspect. This is
somewhat difficult to understand, as it implies that non-sentient
matter has a primitive form of what we call consciousness. The interpretation
also suffers from the combination problem \cite{key-6} i.e. forming
a coherent, single consciousness out of the collection of individual
atoms or neurons of the brain. 
\begin{itemize}
\item Regarding physical matter to be the interface or boundary between
both types of nothing means that matter has both material and non-material
aspects. Therefore, there is a type of property dualism inherent in
the model. This makes sense of the panpsychist view that non-sentient
matter may have a non-material aspect. In the new model, the non-material
aspect is understood to be an essential feature of matter, as it is
required to form the structure of matter. 
\item In regards to the combination problem, while the matter has material
and non-material aspects, it also has a character of substance dualism;
as behind the boundary there is a unified substance (i.e. the material
or non-material nothings). Therefore, this may be how the combination
problem is overcome. 
\end{itemize}
\item \emph{Substance dualism: }In substance dualism it is difficult to
understand where the substance of the non-material consciousness is
located, and how it is tethered to physical matter. For instance,
in the idea that there is a consciousness separate to the physical
brain, it is unexplained how the substance of consciousness is connected
to the physical neurons.
\begin{itemize}
\item If the material and non-material worlds form a boundary of intersection
between them, then it helps to understand in substance dualism where
the non-material substance is located i.e. on one other side of the
boundary, and how it is connected to the material world i.e. it is
intrinsically connected via the boundary of intersection. 
\end{itemize}
\item \emph{The mathematical gap problem: }It is not evident how the non-material
world of consciousness is able to interact with the material world.
If any interactions occur, one might think that they must be represented
in mathematical terms, so as to interface with physical matter which
has a mathematical description. However, any mathematical description
could itself be thought of in material terms, and thus incorporated
as part of the material world. Therefore, this still leaves a divide
between the material and non-material worlds, which does not seem
possible to cross. 
\begin{itemize}
\item The model helps to explain the mathematical gap problem, as there
is no gap. The worlds directly interface, and the motion of matter
involves a co-movement of both sides, to generate a change in the
boundary interface. 
\end{itemize}
\item \emph{Substance vs. property dualism: }No consensus has been reached
on whether substance or property dualism is preferred. Both types
of dualism have unresolved ontological issues. 
\begin{itemize}
\item This new model is a unification of both substance and property dualism.
The substance aspect is evident, because the substance (e.g. in this
case two two types of nothing) lies behind the boundary. Meanwhile,
the property aspect is evident, as the matter is formed by intersection
of both the material and non-material nothings. This may help reconcile
the differences between the substance and property dualist views,
and can perhaps be considered a third type of dualism distinct from
both. 
\end{itemize}
\end{enumerate}

\subsection{Is there anything at all?}

What is traditionally believed to be something e.g. matter, can possibly
be thought of as the intersection of two types of nothing. This means
that there is no ``thingness'' to matter. It is not composed of
any substance. It is a relational structure, not an object in of itself.
The fact that matter might be a relational structure rather than a
substance has been understood by many people historically. 

It is purely a matter of preference whether one wishes to believe
that there is anything `inside' the equations of physics or alternatively
if what's important is the relational structure itself. All that can
ever be deduced from experiments are the `outlines' of the mathematical
symbols defining the equations of motion i.e. the particle behaviour.
But it can never be directly observed what is `inside' (or represented
by) the mathematical symbols i.e. the particles themselves. 

This is evidently connected to the measurement problem, because the
measurement device itself is a part of the physical system. Therefore,
it is impossible to ever directly observe an object. Experiments can
only reveal the relationship of matter (the experimental subject)
to other matter (the measurement apparatus), and so indicate relational
structure not inner substance. 

Therefore, perhaps a conceptual error has been made that there is
any material substance to be found in the first place. Taking a shift
of perspective, what this might be suggesting is that a natural interpretation
is to assume there is no substance, i.e. nothing. Then, what the mathematics
represents is a boundary or interface, which naturally describes relationships,
but does not require an object or inner substance. An interface between
what? An interface between two different forms of nothing. 

More speculatively, this likewise hints that the non-material substance
also may be nothing. It is not necessary to describe any `thing' inside
the conscious experience, only the conscious experience itself, which
may be a relational phenomena. Indeed, this is what many people who
practice meditation or introspection of their cognitive processes
assert. Trying to identify consciousness as a thing, or a real substance,
may share similar difficulties of trying to identify matter as a thing,
or real substance. In both cases, the phenomenon in question can only
be understood in a relational sense, and it is not clear if it can
be identified as a substance empirically in principle. 

\section{The aether}

\subsection{The embedding paradox: }

The notion of having objects existing in isolation in empty space
is somewhat paradoxical. When taking an abstract, relational view
of spacetime, there is the concept that the spatial location is just
a relational construct. For instance, the frame of reference (e.g.
stationary, moving, rotating) is relational, as is the conformal size
of the space and distance between objects. 

Now, if the background space is abstract and relational rather than
static and concrete, then there is a problem of understanding how
a real physical object can be embedded in this abstract space. This
is similar to the problem of understanding how a closed-finite universe
can be embedded in an external abstract void. In both cases, it is
difficult to understand how something real can have a boundary with
something abstract. 

These problems are alleviated if the physical matter is not a `something'
but is a complex structure present in the interface of two types of
nothing. Then there is no need to place the object inside the abstract
space, since the object naturally arises out of it. Therefore we have
returned to the concept of the aether. If physical matter really is
composed of this boundary, then in the boundary is analogous to the
aether. 

\subsection{Mach's principle: }

Describing the boundary of intersection as a type of aether might
help understand the origin of inertia, which is a mystery that can
be explored through Mach's principle. It is clear that the paradox
of Mach's principle is still present, despite general relativity making
an attempt toward the problem. The thought experiment of Mach's principle
is this: 
\begin{quote}
Imagine a rotating object in a space surrounded by other mass (for
instance the distant stars). Due to inertia it experiences a centrifugal
force. For example, spinning in a circle will cause your arms to stretch
out. However, if one imagines this same object spinning in empty space,
in the rotating frame of reference there is no motion. Due to the
absence of other external objects, this rotating frame of reference
can be regarded as the stationary reference frame, and so presumably
no centrifugal forces appear. The question is, what physics distinguishes
the two scenarios? 
\end{quote}
The answer provided by general relativity is that the metric tensor
$g_{\mu\nu}$ which encodes the rotating frame of reference (and hence
the centrifugal forces) is determined by the stress energy tensor
$T_{\mu\nu}$ which contains the mass-energy of the distant stars.
This was Einstein's attempted solution to the problem. However, there
are two peculiarities in this viewpoint.
\begin{enumerate}
\item The centrifugal forces occur immediately. There is no delay, or propagation
of actual gravitons or gravitational field from the distant location
to the spinning object. 
\item Moreover, if the gravitational field or gravitons are involved, one
might think that they share the same problem of having to be grounded
in a reference frame. Therefore, it is not clear that one truly escapes
the problem or merely shifts it. 
\end{enumerate}
Therefore, general relativity does not appear to fully solve the problem.
The problem is that without an aether, there is a lack of a preferred
background reference frame from which the concept of inertia emerges.
However, if matter is understood to be comprised of the boundary between
two types of nothing, then the aether concept is restored. This provides
a background reference frame which can possibly explain inertia. 

\subsection{Holographic principle }

This proposal may have a connection to the holographic principle,
as it suggests that matter exists as a structure on a two-dimensional
sheet which is the boundary of intersection. Further research may
be needed to investigate this connection. 

\section{Causation}

\noindent The proposed model provides a new understanding of physical
causation, as the motion of matter is described by motion of the boundary
interface. When objects move in space, it involves a conjoined motion
of both the material and non-material sides of the boundary. The boundary
might move like an undulating wave, which involves a synchronized
movement of both sides. This is important for several reasons, which
will be discussed in this section. 

\subsection{Epiphenomenalism }

Just as there is a problem of understanding how the non-material consciousness
can affect the material world, there is the converse problem of understanding
how the material world can drive the time-evolution of the non-material
consciousness. For instance, suppose consciousness is a substance.
In this case, the matter undergoes time evolution, but how does the
consciousness update and undergo time evolution? 
\begin{enumerate}
\item One possibility is the consciousness is continuously generated from
the underlying matter. In this case, there is a problem, because it
is not clear how matter interfaces with the non-material world so
as to generate this consciousness. This is like the matter of the
brain as a roll of film, and the consciousness is being played as
a projection of this film upon a screen. Yet there is no known projector
mechanism involved; the hypothetical projector mechanism is susceptible
to the mathematical gap problem. 
\item Another possibility is there are causal influences which cross from
the material world to the non-material world, to inform the time-evolution
of the non-material consciousness. This likewise has the mathematical
gap problem. It is not clear what these causal influences could be
ontologically, as they must cross the divide between the material
and non-material worlds. 
\end{enumerate}
This problem is solved by eliminating the gap between material and
non-material worlds, by having the two worlds directly interface one
another. This is an important point, because while it is contentious
that consciousness can play a causal role in the physical dynamics
of the material world, there is no doubt that the material world must
have a causal influence over the non-material world (otherwise the
physical brain states would become dissociated from the states of
consciousness). 

There must be a causal mechanism permanently locking these two things
together; the physical brain state and the state of consciousness.
In the model of two intersecting forms of nothing, these two things
are intrinsically locked together at a fundamental level. 

It should be noted that property dualism also solves this problem
to a degree. If the individual particles have material and non-material
aspects, then the two worlds are also locked together. However, this
raises a new problem of how a single, unified conscious experience
can arise out of the individual elements (i.e. the combination problem).
In our new model, the combination problem is overcome.

\subsection{Interactionism}

As indicated above, even in the epiphenomenalist view it is difficult
to explain consciousness. This problem is compounded when considering
the evidence against epiphenomenalism. This evidence includes i) the
philosophical zombie argument ii) an information argument, that information
arising within conscious experience can be transferred to the physical
world \cite{key-5} and iii) an evolutionary argument \cite{key-7}.
If these arguments are correct (and we believe their case is strong),
then it becomes clear that a mechanism is needed to link the material
and non-material worlds. This mechanism is provided by the new model,
where matter is formed as an intersection of the material and non-material
worlds. 

\subsection{Materialistic determinism vs. idealistic determinism: A matter of
perspective? }

In regards to determinism and free-will, it may be the case that the
motion of the matter is not solely determined by the material aspect,
nor solely by the non-material aspect. Rather both aspects work in
unison to create the motion. This allows room for conscious experience
to play a causal role. However, it is not an active form of causation,
it is more passive in nature as the non-material aspect is being shaped
by the motion simultaneously as it occurs.

We believe that this may bridge the divide between deterministic materialism
and the view that consciousness plays a causal role. In fact, from
one point of view deterministic material time-evolution may be true.
But from another point of view, the non-material world may be completely
governing the motion i.e. idealism.

This is because physical matter is composed by both sides of the boundary,
the material and non-material sides. Materialistic determinism views
the motion of matter purely from the material side of the boundary.
Then viewing from the opposite side of the boundary, perhaps there
is likewise a perspective where the non-material world completely
describes the motion. This is another way by which consciousness could
be understood to affect the material world. However, in reality, both
sides may co-cause the motion.

The problem is not that deterministic materialism is wrong. A deterministic
materialist explanation is always possible looking backward at the
time-evolution of the physical system; but it comes at the cost of
creating fine-tuning of the microcausal degrees of freedom \cite{key-5}.
A pure materialistic determinism is unlikely in the forward-sense
of the time evolution, as the probabilities for the particle distribution
to go down specific branches of the wavefunction are affected by non-material
phenomena (e.g. conscious experience). The same may be true from the
perspective of the non-material side; it may have a retrospective
deterministic explanation, but be fine-tuned in the initial degrees
of freedom when looking backward at the time-evolution. Only by considering
both the material and non-material aspects simultaneously can the
fine-tuning be removed. 

\section{Conclusion }

\noindent We have proposed that matter is comprised from the boundary
of intersection between two types of nothing; the nothing of the material
world, and nothing of the non-material world respectively. Assuming
that void states are impossible (e.g. the total void, material void,
and non-material void), then the world either contains something,
or two types of nothing. 

If it contains two types of nothing, then this is not equivalent to
a blank empty space. Taking as an ansatz that the two types of nothing
a) share the same physical space and b) are mutually exclusive and
cannot overlap in spatial location, then they must have a boundary
of intersection between them. Due to reasons of arbitrariness, this
boundary of intersection is likely unstable, and spontaneously forms
complex structures, thus providing a plausible mechanism for matter
to emerge from the primordial space.

The reality of the physical world is that it likely does have two
aspects, the material and non-material worlds. This is shown by the
hard problem of consciousness, which demonstrates the existence of
the non-material world. Taking the dualist view, this is typically
regarded as an expression of two types of something. However, it is
possible to alternatively conceive of two types of nothing. 

This new model helps to explain dualism, as it unifies property and
substance dualism into a single approach that has aspects of both.
The model may also have significance as a way to describe physical
causation. Since both sides of the boundary are co-moving and inherently
conjoined, the model can explain how causal influences are able to
cross from the material side to the non-material side and vice-versa,
eliminating the mathematical gap problem of dualism. The model furthermore
can be thought of as a return to the aether concept, thus providing
a possible means to explain the origins of inertia and Mach's principle.
The model's status in relation to the holographic principle remains
unexplored. 

\subsection*{Statement of originality}

All work is original research and is the sole research contribution
of the author. 

\subsection*{Conflicts of interest: }

No conflicts of interest. 

\subsection*{A.I.: }

No artificial intelligence was used in the production or editing of
this work.

\subsection*{Copyright notice: }

Copyright{\small{} }{\footnotesize{}© }2024 Adam Brownstein under
the terms of arXiv.org perpetual, non-exclusive license 1.0.

\end{document}